# A Social Network Analysis of JWST General Observer Programs: The Emergence of a Decentralized Heterarchy

Christopher Williams
ESSCA School of Management
55 Quai Alphonse le Gallo
92513 Boulogne-Billancourt
France

Email: christopher.williams@essca.fr

## ABSTRACT

The development of proposals to execute programs on space telescopes involves networks of astronomers coalescing around ideas and plans for observations. When aggregated together, these program-level networks allow a collective structure of the overall social network of astronomers using a telescope to be created and analysed. We do this using program-level investigator data over the first five cycles of accepted General Observer (GO) programs on the James Webb Space Telescope (JWST). The aggregate network contains n=5252 unique astronomers (nodes) with 144740 connections (edges) between them based on their program-level participation. We apply a modularity class coefficient to visualize the sub-communities that evolved within the aggregate network. Ten dominant sub-communities emerge that are shown to correlate at various levels with the JWST scientific categories as defined by the Space Telescope Science Institute (STScI). These sub-communities vary in size as well as diversity of countries and institutions represented. Analysis of the research interests of investigators within sub-communities reveals that separation along the long axis of the graph is associated with the scale of science performed at community level (AU-scale vs kpc-Gpc scale). While the social network structures based on institutional affiliation and country of institution are highly centralized, the aggregate network at the investigator-level is highly heterarchical and decentralized. It appears to be supportive of cross-disciplinary interaction, and institutionalized integration between planetary-system and cosmic structure science that is wide and redundant, rather than mediated by a small set of critical brokers. Results have implications for access policy (time allocation processes on telescopes leaving behind a legacy of heterarchical social networks) and access strategy (astronomers' need to access networks before accessing telescopes).

**Key words:** James Webb Space Telescope (JWST); social network analysis (SNA); scientific categories; scale.

## 1. Introduction

Many astronomers, like scientists in numerous other fields, work within international networks in pursuit of scientific goals. These networks are social networks, defined as "… a set of socially-relevant nodes connected by one or more relations." (Marin and Wellman, 2011, p.11). The phenomenon of international scientific networks in modern astronomy is partly a consequence of developments in IT, communications and database technology. These enable meaningful and productive social relations to form between scientists from different locations while allowing access to distributed resources for scientific work (Andernach et al., 1994). The phenomenon is also a natural consequence of the need to pool expertise and knowledge to solve complex problems; knowledge that exceeds the capabilities of any one single investigator (Wuchty et al., 2007). The need for cross-disciplinary networking in astronomy is strong; the Astro2020 Decadal Survey (NASEM, 2021) suggests that progress in its priority science areas will increasingly depend on collaboration across scientific communities and observational domains. These collaborations are embedded within international social networks of astronomers.

In the field of astronomy, networks of investigators coalesce around specific programs to use terrestrial and space telescopes to conduct observations and collect light. Astronomers are not tied to one network or one telescope; they can seek to pursue their research goals by participating in different networks, using different telescopes, with different instruments over time. While there is a large amount of collaboration in networks to conduct observations through programs on telescopes, there is also keen competition to gain access, this having become more intense for the most powerful and in-demand assets (Rao, 2024; Williams, 2025).

Much of the research on networks in astronomy has not looked at the proposal-, or program-level networks linked to telescope usage. Instead, it has tended to focus on networks at the 'output-side' of scientific endeavour, namely author-based, publication-based and citation-based networks (Espinosa-Rada et al., 2024; Espinosa-Rada, 2026; Heidler, 2011) or telescope bibliographies (D'Abrusco et al., 2023). We argue that a more complete understanding of the phenomenon of network formation in astronomy can be achieved through social network analysis (SNA) of collaboration networks at the proposal- or program-level on specific telescopes. In contrast to author-based networks, this represents the 'input-side' of the phenomenon and is important because it can capture the formation and nature of astronomy networks before discoveries are made or published. It represents a point in time where scientific vision, curiosity and intention take greater prominence over maximizing publications, citations and impact.

Terrestrial and space telescopes on which programs are run are nested within institutions and countries. In some instances, the institutions that build and operate telescopes are based in one country (e.g., FAST, China). In others, one country may be a dominant player alongside other international partners (e.g., Hubble, JWST; USA as lead country). In other situations, the collaborations to build and operate are highly multinational (e.g., ALMA, EST, GMT). Individual investigators participating in proposal networks can be drawn from a much wider set of institutions and countries for international (dominant player) (e.g., Hubble, JWST) and multinational (e.g., ALMA) telescopes. General Observer (GO) programs through annual open calls allow participation by investigators from institutions and countries that were not involved in the design, building and operating of the telescope. This makes program-level networks interesting to understand as it is possible to analyse collaborative activity in the broader field, as well as across levels: country, institution, and individual astronomer. At each level graphs can be constructed with nodes (countries, institutions, or astronomers) and edges (between countries, between institutions, and between astronomers respectively). Expectations for country and institutional networks may be influenced by the fact that country and institutional involvement in the funding, building and operating of the asset can bias access (Langford and Langford, 2000; Malkov et al., 2011; McCray, 2000). These core countries and institutions are associated with centres of knowledge and technical expertise related to instrumentation and telescope use (Shrum et al., 2007). Consequently, we might reasonably expect the 'input-side' networks for telescope programs at country and institution level to be highly centralized (e.g., Hubble and JWST centralized around the US and dominant US institutions). This is reinforced by research showing astronomy to be characterized by collaboration networks linking institutions internationally, with prominent hubs and leading institutions occupying central positions within the global collaboration structure (Zhou et al., 2014).

Expectations for investigator-level networks, on the other hand, are arguably more nuanced. At this level, scientific sub-categories within which astronomers work take a more prominent role (Williams, 2025). For example, STScI defines eight scientific categories for JWST, with definitions and labels for some categories changing over the cycles. Categories can be grouped into solar system and local group, galactic, and extra-galactic super-categories (Williams, 2025). Investigator-level networks may also be influenced by the instrument ecosystem related to a given program. There may be tighter relational ties between members of a given instrument ecosystem because of knowledge and experience of technical specifications, calibration, data reduction and processing, and software tools and training needed for analysis of data from a given instrument. The four scientific instruments on JWST were developed by different institutional consortia (Rieke et al., 2005), potentially creating separate clusters of technical expertise around each instrument. A third reason why investigator-level networks may differ to institution and country networks is emerging policy designed to broaden participation among underrepresented groups and countries with limited astronomical infrastructure. The increasing internationalization of astronomy is reflected in the growth of the International Astronomical Union, whose national membership expanded from eight member countries in the early twentieth century to 85 national members by 2022, while individual membership exceeded 13,000 astronomers worldwide in 2025 (IAU, 2026). This professional astronomer community is distributed across many countries and institutions of the world, extending far beyond the small group of powerful countries and institutions that lead on telescope funding, design, development and operations. For these reasons, we may expect investigator-level social networks captured through telescope use in astronomy to be less centralized than those at the country and institution levels.

To the best of our knowledge, no study has been conducted to date using SNA on program level investigator data for JWST. At the time of writing, JWST was the most powerful and expensive telescope ever commissioned and among the most in-demand scientific facilities ever constructed (Rao, 2024). Proposal success rates declined to below 10% by Cycle 5. We apply SNA to program level investigator data over the first five years of General Observer (GO) programs (method described below) to answer the following related research questions:

> *How do aggregate country, institution, and investigator level networks differ on the 'input-side' to JWST usage (Question 1) and how does the investigator network shape and integrate JWST scientific communities (Question 2)?*

## 2. Results

<u>*Question 1: Differences between country, institution, and investigator level networks*</u>

Figure 1 shows the network graphs at country (top) and institution (bottom) level respectively.

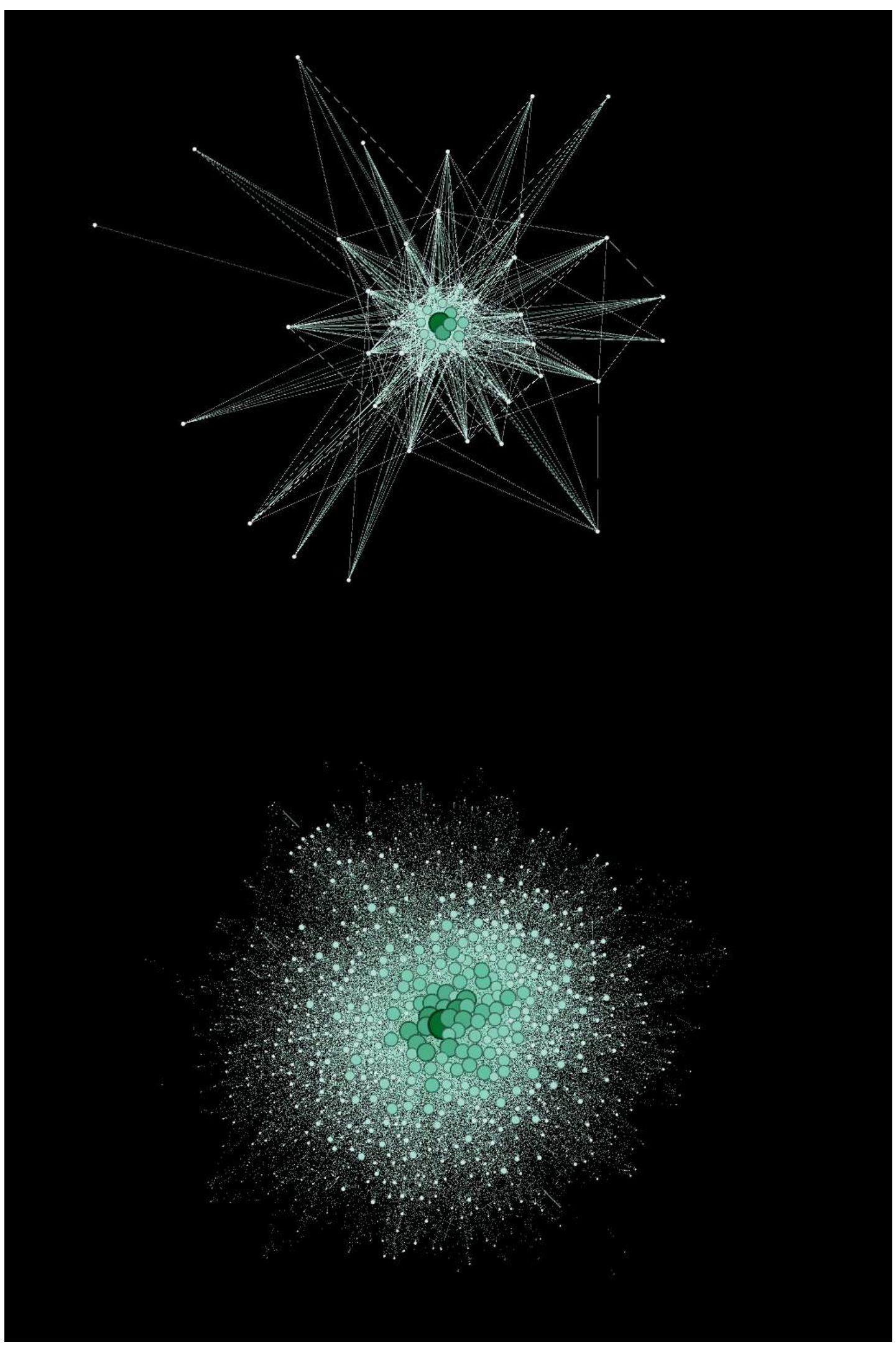

Country {Nodes = 52; Edges = 625; Ave Degree = 24.038; Modularity = 0.004; Ave Clustering Coeff. = 0.832; Density = 0.471; Degree centralization = 0.529}

Institutions {Nodes = 829; Edges = 33380; Ave Degree = 80.531; Modularity = 0.158; Ave Clustering Coeff. = 0.735; Density = 0.097; Degree centralization = 0.697}

*Figure 1. Graphs at country level (top) and institution level (bottom)*

At the country level (top graph in Figure 1), there are 52 nodes (countries) with 625 connections between them. The US has the highest number of connections to other countries (50, compared to an average of 24.038 across the network), and assumes a central position in the graph (dark green node). The countries immediately surrounding the US in the graph are UK, Germany, and France, followed closely by Switzerland, Denmark, Japan, Australia, Netherlands, Canada, Italy and Spain. There is a high degree centralization ($C_D = 0.529$) suggesting a small number of countries account for a large amount of collaboration. Countries at the periphery of the network include Cyprus, Singapore and Croatia (selected examples). The network is a very dense, highly interconnected network with almost no discernible community structure (low modularity, no meaningful sub-groups at country level). On average, each country is connected to nearly half of all the other countries in the network (ave. degree – 24.038). The country network is a single, integrated collaboration community, rather than a collection of sub-communities. In this type of network, information, resources, and knowledge can flow quickly and efficiently.

At the institution level (bottom graph in Figure 1), there are 829 nodes (institutions) with 33380 connections between them. STScI has the highest degree (656 connections to other institutions; note ave. degree is 80.531) and assumes the central position (dark green node). $C_D$ is 0.697, higher than for the country graph. The institutions immediately surrounding STScI include Johns Hopkins University, Max Planck Institute for Astronomy, California Institute of Technology, University of Leiden, NASA Goddard Space Flight Center, University of Arizona (selected). At the periphery of the graph, we see institutions such as Tokyo Metropolitan University, University of Hyogo, Technical University Dresden, and Worcester State University (selected examples). This is a larger, more centralized, but less dense network compared with the country graph. There is weak-to-moderate community structure (the modularity coefficient (Q) increases over the country level graph, but is still relatively low), indicating that scientific collaboration is organized around a relatively small set of leading institutions.

Figure 2 below shows the graph at individual investigator level, before (top) and after (bottom) modularity class coding is applied.

At this level, there are 5252 nodes (investigators), with 144740 connections between them. The modularity coefficient (Q) increases compared to country and institution graphs, while average degree and average clustering coefficients are within the range for country and institution levels. When we consider Figure 2 (top), there is no one single central node with a degree close to the total number of nodes in graph (as we see at country and institution levels). The ForceAtlas2 algorithm does not create a centralized graph as it does for country and institution levels. Instead, it creates a rather heterarchical (Stark, 2009), dispersed and scattered graph with multiple centres. Degree centralization ($C_D = 0.128$) is low compared to country ($C_D = 0.529$) and institution graphs ($C_D = 0.697$). The most prominent cluster appears to be towards the right of the graph, with various clusters slightly to its left, and another cluster to the very left of the graph. The increase of Q to 0.644 suggests the investigator network is composed of distinct collaboration groups rather than one integrated community. Global density is low (i.e., sparse) while local density is high; most links occur *within* communities, with relatively fewer links connecting different communities. This is indicative of a sub-field structure within the aggregate investigator network; researchers collaborate intensely within their own communities, and those communities are themselves highly cohesive.

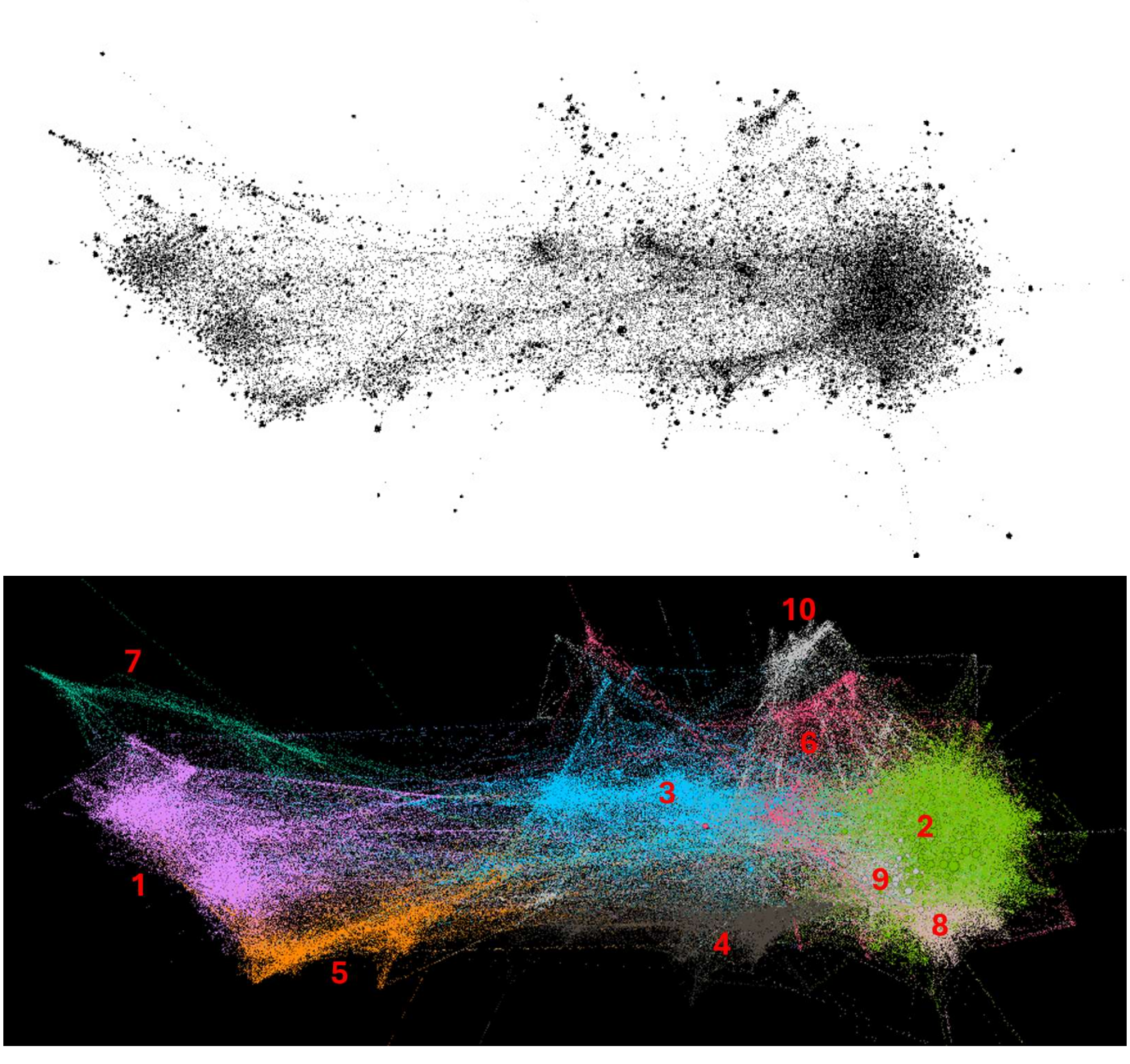

Investigator {Nodes = 5252; Edges = 144740; Ave Degree = 55.118; Modularity = 0.644; Ave Clustering Coeff. = 0.759; Density = 0.01; Degree centralization = 0.128}

*Figure 2. Graphs at investigator level – before (top) and after (bottom) modularity class coding*

The bottom graph in Figure 2 shows the network after applying the modularity class as the partition attribute for node colour. There are 10 dominant, distinct communities that make up 95.42% of the nodes in the graph. These are labelled in Figure 2 in descending order of community size. There are a further 10 communities that make up the remaining 5.48%, with contributions of between 0.04% (3 nodes) to 1.69% (89 nodes) of the graph. We apply a cut-off at the top 10 clusters, to focus on the main sub-fields and minimize noise. While ForceAtlas2 discriminates between communities 2, 8 and 9 on the right-hand side, these communities are spatially closely related. Table 1 below shows the legend for identified communities along with their distribution and interpreted scientific sub-field. The correlation of the SNA-derived communities with STScI scientific categories is made through program affiliation of investigators and yields high correlations – albeit to varying levels - with the string matches shown (Table 1).

| ID | Distr. (%) | Nodes | Community interpretation | Match with STScI scientific category label | STScI category largest match | Science category diversity | Country diversity | Institution diversity |
|---|---|---|---|---|---|---|---|---|
| 1 | 20.64 | 1084 | Exoplanets | 83.2% - “exoplanets” | Exoplanets and Exoplanet Formation (49.7%) | 4.04 | 3.37 | 97.6 |
| 2 | 18.58 | 976 | Galaxies and the inter-galactic medium | 93.1% - “galaxies” | Galaxies (47.7%) | 4.60 | 4.03 | 102.5 |
| 3 | 13.02 | 684 | Stars and stellar systems | 86.9% - “galaxies” or “stellar” | Stellar Populations and the Interstellar Medium (26.0%) | 6.92 | 3.86 | 92.0 |
| 4 | 9.41 | 494 | Stellar-evolution, supernova-remnants, and dust-cycles | 86.4% - “stellar” or “dust” | Stellar Physics and Stellar Types (43.2%) | 5.13 | 2.63 | 78.9 |
| 5 | 8.72 | 458 | Planet formation and protoplanetary disks | 98.6% - “stellar” or “dust” or “exoplanet” | Exoplanets and Exoplanet Formation (27.0%) | 5.89 | 5.98 | 97.4 |
| 6 | 8.05 | 423 | Galaxies and galaxy clusters | 84.7% - “galaxies” | Supermassive Black Holes and Active Galaxies (60.3%) | 3.77 | 5.01 | 95.8 |
| 7 | 5.05 | 265 | Solar system | 95.9% “solar system astronomy” | Solar System Astronomy (95.9%) | 1.20 | 2.27 | 42.7 |
| 8 | 4.53 | 238 | Galaxies / early universe | 97.8% - “galaxies” or “large scale structure” or “intergalactic medium” | Supermassive Black Holes and Active Galaxies (51.7%) | 4.38 | 4.87 | 52.3 |
| 9 | 3.87 | 203 | Galaxy evolution and cosmic history | 89.4% - “galaxies” | Galaxies (36.7%) | 5.71 | 6.77 | 23.6 |
| 10 | 2.65 | 139 | Time domain and high-energy transients | 98.0% - “galaxies” or “stellar” | Stars and Stellar Populations (37.6%) | 3.98 | 7.14 | 41.4 |

Note: Diversity values are Inverse Simpson Index by category; higher values denote greater diversity

*Table 1. Distribution of communities with scientific focus and community diversity*

The Inverse Simpson Index of diversity for STScI scientific categories shows community #7 (solar system) to be the most concentrated, and community #3 (stars and stellar systems), the most diverse in terms of STScI scientific categories within the community. Table 1 also shows the diversity scores for institutions of nodes within each community, as well as diversity in terms of the countries of the institutions for nodes within communities. The most concentrated community in terms of country diversity is again community #7 (solar system) – heavily dominated by the US - and the most diverse is community #10 (time domain and high-energy transients). The most concentrated community in terms of institutions is community #9 (galaxy evolution and cosmic history) – dominated by the University of Cambridge and the University of Arizona - while the most diverse is community #2 (galaxies and the inter-galactic medium). Bi-variate correlations between community size (i.e., node count) and the diversity variables yields Pearson's $r$ of 0.19 for scientific category diversity, -0.47 for country diversity, and 0.79 for institution diversity. This is depicted in the drop line graph in Figure 3 below which shows the community IDs in magenta and has a y-axis which is a z score of the diversity values (mean 0, standard deviation 1) to allow comparison. Smaller sized JWST communities tend to have a higher country diversity and lower institution diversity, while larger communities tend to have a higher institution diversity and lower country diversity. Solar system astronomy is an outlier in the sense that it is the only community with lower diversity than the mean on all three indices. Community #3 (stars and stellar systems) is notable for having the highest science category diversity and an above average institution diversity.

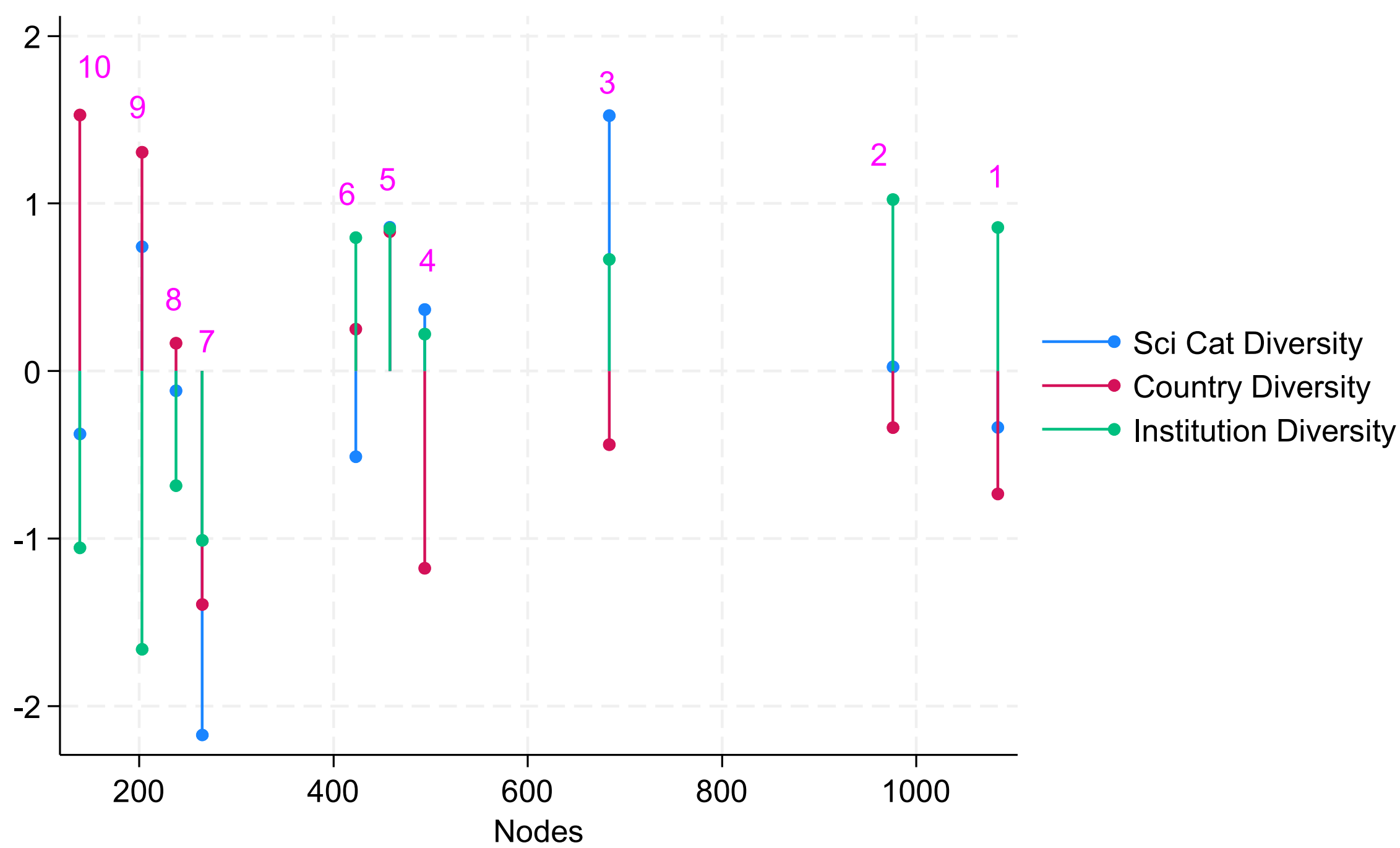

Note: y-axis is normalized diversity (mean = 0); Line labels in magenta are community IDs from largest (1) to smallest (10) (see Table 1)

*Figure 3. Relationship between community size and diversity*

*Question 2: How the investigator network shapes and integrates JWST scientific communities*

Figure 4 shows an interpretation of the dimensions of the investigator graph. As the graph is longer wide than it is high, we first consider a notional x-axis along the horizontal length of the graph. We find that nodes towards the left of the graph tend to belong to research communities undertaking science on AU-scale objects, typically 0.01 – 10,000 AU. The three communities of solar system (7 - turquoise), exoplanets (1 - pink) and planet formation and protoplanetary disks (5 - orange) (Figure 2 and Table 1) contain 1807 investigator nodes; these three communities align with AU-scale science. Investigators in these communities are more likely to be involved in programs that characterize constituents and physical states that are directly observable, this being particularly relevant for solar system astronomy (including comets, asteroids, icy satellites); exoplanet atmospheres and compositions; dust, gas, ices, and molecular inventories. In terms of instrument use, these communities rely more on NIRISS (e.g., exoplanet spectroscopy) compared to other communities and are more likely to use coronagraphy to suppress starlight and observe planets. They are also heavily associated with JWST's Time-Series Observation (TSO) mode.

The remaining five communities in the centre and right-hand side of the graph are made up of 3157 investigator nodes. These investigators are involved in programs that identify formation paths, evolutionary timelines and test models of galaxy, black-hole, and stellar evolution. Their work examines large scale cosmic structure at kpc–Gpc scales. Many of the researchers on this side are not just studying individual galaxies (kpc scales), but also galaxy populations across the observable universe, cosmic reionization, large-scale structure, galaxy clustering, and cosmological evolution and the first billion years of the Universe. They are more likely to use multi-object spectroscopy and survey capabilities compared to the AU-scale communities.

Investigators towards the AU-scale end have a proclivity to study objects that are not inherently redshift dependent. Those towards the kpc–Gpc side are strongly redshift dependent. Note: Various tests of community group practice based on modularity class coding along a notional vertical y-axis do not reveal any clear and consistent differences between those at the bottom of the graph and those at the top. This will be examined further in future work.

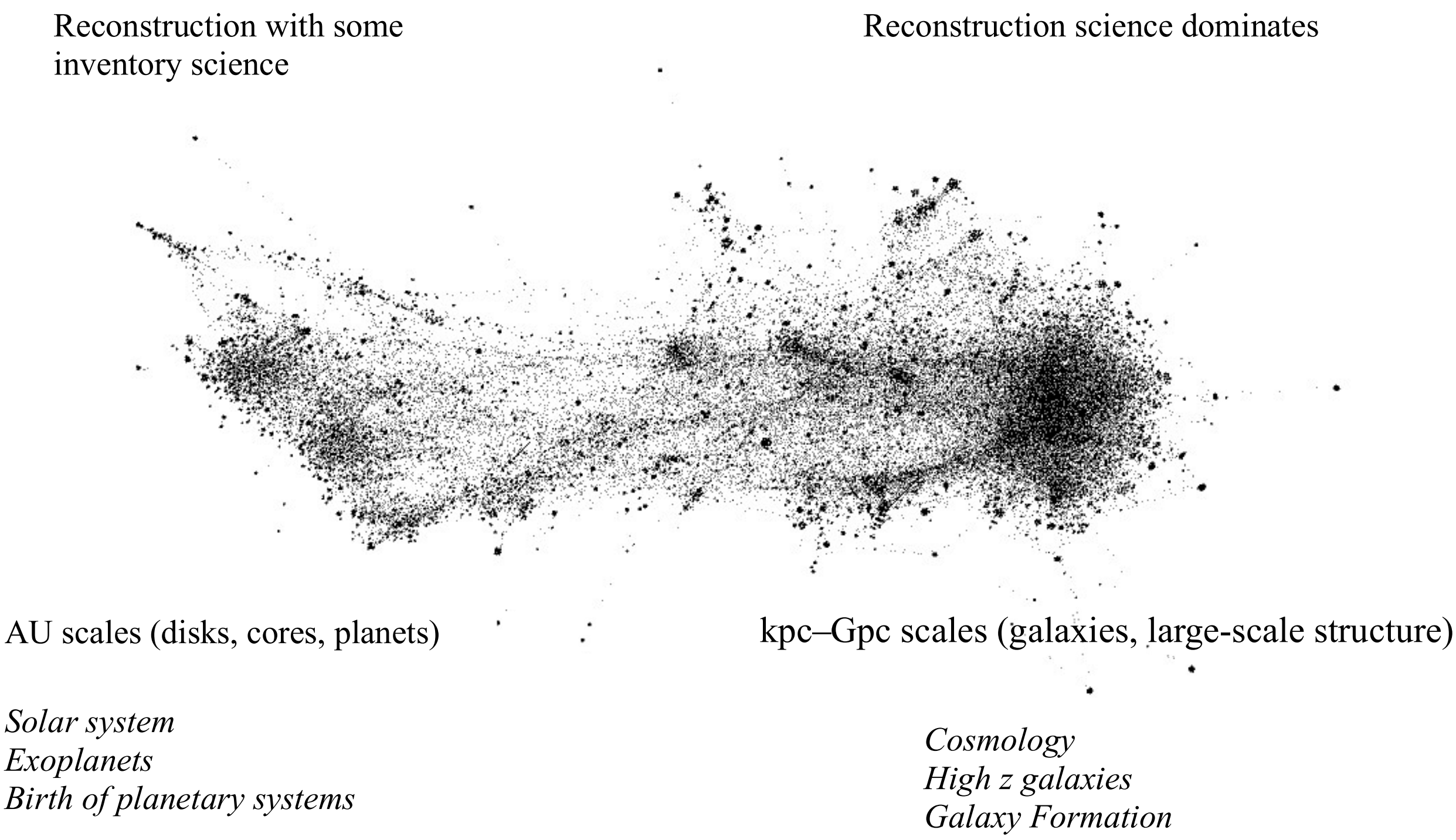

*Figure 4. Interpretation of the JWST investigator graph*

Examples of GO programs linked to nodes in the AU-scale side of the graph are:

- GO1846: A Search for Signatures of Volcanism and Geodynamics on the Hot Rocky Exoplanet LHS 3844b
- GO5967: Exploring the desert: Thermal characterization of an exposed planetary core
- GO10488: Uranus Solstice Explorer: Temporal Change in the Atmosphere, Ionosphere, and Ring-Moon System

Examples of GO programs linked to nodes in the kpc-Gpc side of the graph are:

- GO2659: Beasts in the Bubbles: Characterizing ultra-luminous galaxies at Cosmic Dawn
- GO5791: The Crucible of Planet Formation - Protoplanetary Disks in the Extreme Environment of Trumpler 14
- GO12707: LensTrove: Unlocking the Cosmological Power of Cluster-lensed Supernovae

All communities in the network have a *reconstruction* quality, i.e., nodes linked to programs that seek to explain the origins and evolution of the universe and answer a key question: How did it become the way it is? However, those towards the left of the network (i.e., AU-scale) have at least some emphases on building census of astronomical objects and their properties. The central question in this type of science is: what is there? We note that inventory science is more prevalent within the AU-scale communities compared to the kpc-Gpc communities.

Examination of the emergent investigator graph reveals a small number of nodes that appear to connect AU-scale communities with kpc–Gpc scale communities. In social network terms, these nodes can be considered *bridging nodes*: links between different groups, clusters, or communities, facilitating the flow of information, resources, and knowledge across the network. Three examples are shown in Figure 5.

Bridging nodes are particularly interesting in a large scientific community because of their potentially powerful positions as brokers of information, resources, and knowledge. They may also act as gatekeepers between communities: controlling, filtering, facilitating, or restricting the flow of information, resources, and knowledge and being valuable for different communities. Not every bridging node will be a gatekeeper in this sense, but they can have an influential role – and therefore be of high value to science – influencing exchanges across remote parts of the network and even creating opportunities for new proposals by linking different areas of expertise. We note a small number of such nodes in the JWST investigator network that are positioned on the graph in a 'middle ground' (i.e., not positioned at the centre of any dominant community). The examples in Figure 5 fit this profile; they have visible edges with AU-scale communities and kpc-Gpc scale communities. These nodes have low clustering coefficients compared to the average for the whole graph (0.759). 126 out of 5252 nodes have a clustering coefficient < 0.25 and, of these, only a handful (c. 10) appear to link AU-scale and kpc-Gpc scale science. 44 out of 5252 have a clustering coefficient < 0.20.

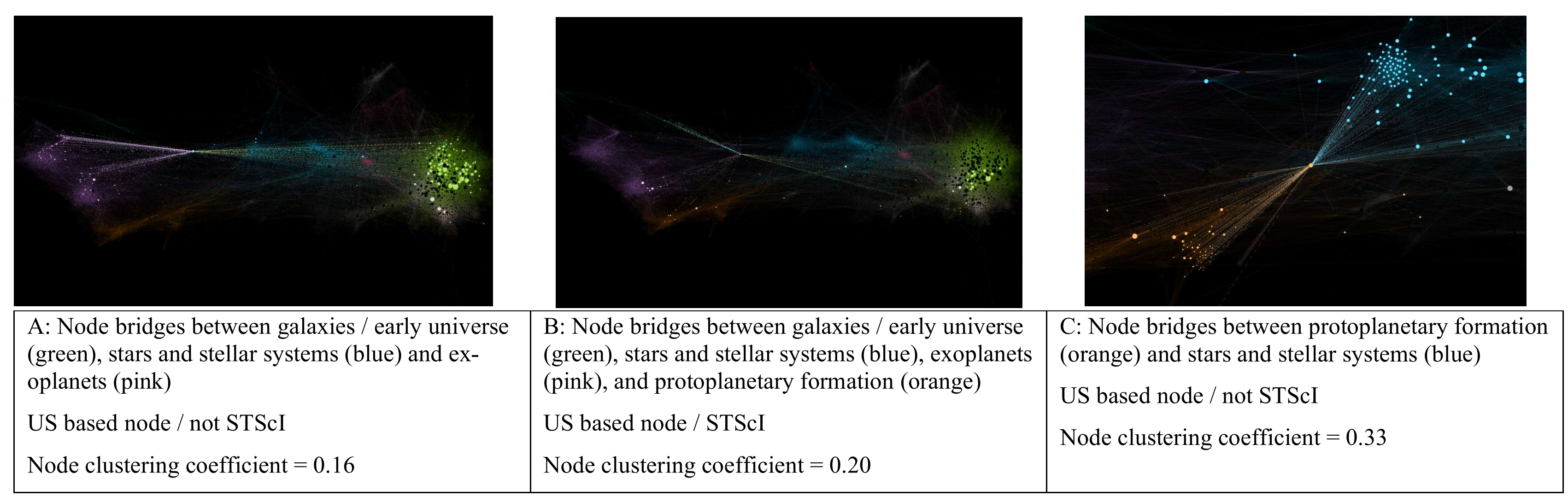

| A: Node bridges between galaxies / early universe (green), stars and stellar systems (blue) and ex-oplanets (pink)<br><br>US based node / not STScI<br><br>Node clustering coefficient = 0.16 | B: Node bridges between galaxies / early universe (green), stars and stellar systems (blue), exoplanets (pink), and protoplanetary formation (orange)<br><br>US based node / STScI<br><br>Node clustering coefficient = 0.20 | C: Node bridges between protoplanetary formation (orange) and stars and stellar systems (blue)<br><br>US based node / not STScI<br><br>Node clustering coefficient = 0.33 |
|---|---|---|

*Figure 5. Examples of bridging nodes in the investigator network*

Figure 6 shows the investigator graph, colour coded based on the modularity class but filtered for nodes with clustering coefficients < 0.20 (top) and < 0.25 (bottom). Various tests to remove bridging nodes and re-run graph statistics reveal no meaningful change in the structural characteristics of the graph. For instance, removal of nodes with clustering coefficients < 0.2 results in a graph with a change of average degree from 55.118 to 54.654, a change of modularity coefficient from 0.646 to 0.643, and a change of average clustering coefficient from 0.759 to 0.760. Figure 6 shows how a small increase in clustering coefficient as a graph filter results in a large increase in the number of edges between AU-scale and kpc-Gpc scale communities. These results suggest bridging nodes are not structural bottlenecks and their roles have been internalized by the overall network.

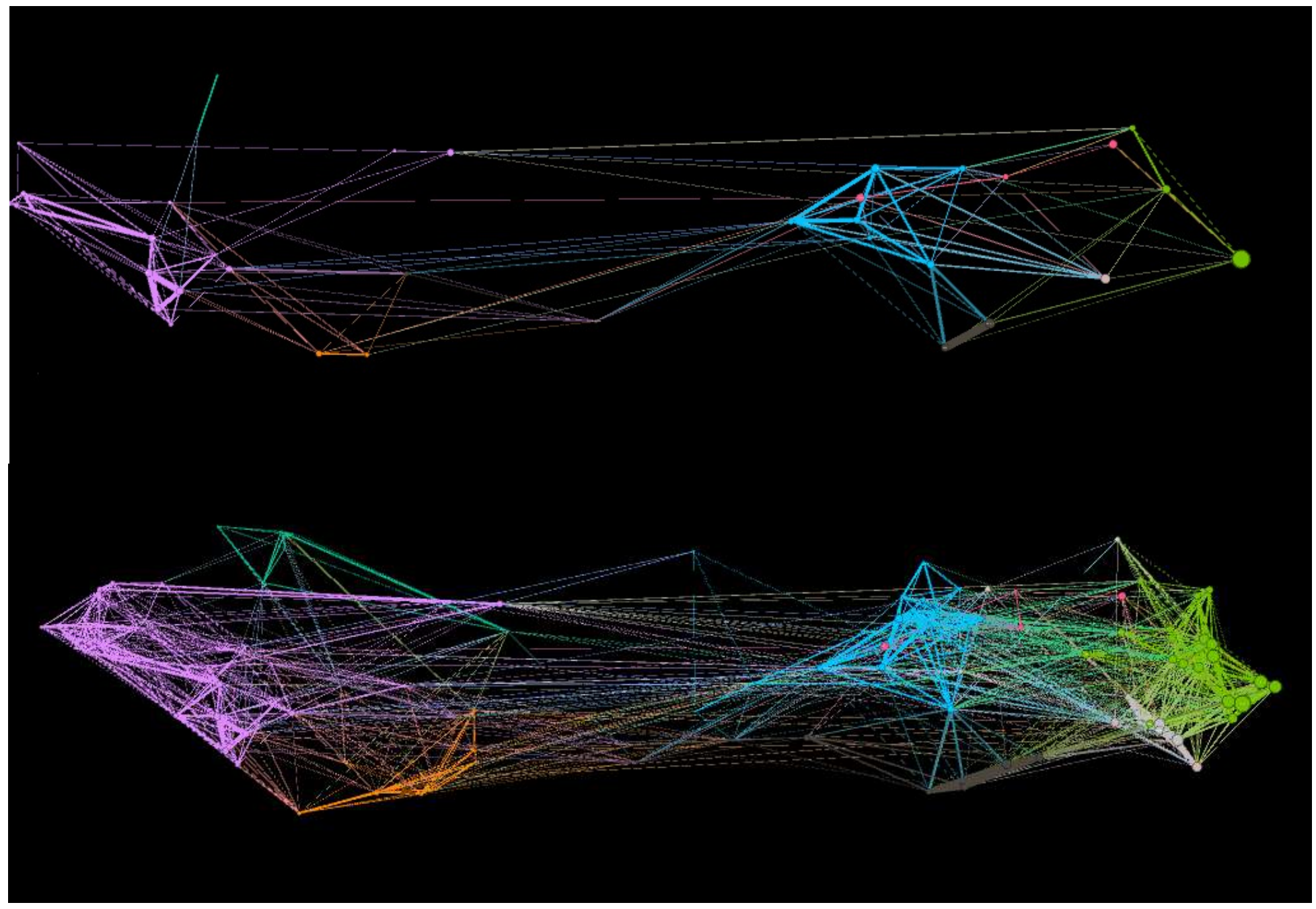

*Figure 6. Bridging nodes based on clustering coefficients (<0.20, top) and (<0.25, bottom)*

## 3. Discussion

The differences between country and institution graphs on the one hand, and the investigator graph on the other, are broadly as expected. Country and institution networks are more centralized than the investigator level, although the institution network exhibits a higher degree of centralization than the country network (0.697 versus 0.529). This suggests collaborative activity on JWST can be seen as being organized more strongly around a small number of hub institutions than around a small number of hub countries. We see no modularity at the country level, weak modularity at the institution level, but high modularity at the investigator level. Again, this follows intuition, although it has not been demonstrated to date on JWST program data. The communities that emerge at the investigator level vary in terms of their prominence and diversity. Over 50% of the investigator network nodes are accounted for by three communities: exoplanets (20.64%), galaxies and the inter-galactic medium (18.58%), and stars and stellar systems (13.02%). While there is a high correlation between the SNA derived communities of nodes as demarcated by modularity class and STScI science categories of programs, there is some variance in the level of the matches (83.2% - 98.0%), indicating most communities contain nodes operating across neighbouring scientific categories (higher related diversification within the community), while others (notably, solar system astronomy), are undiversified. The communities also vary according to country and institution diversity within them. Larger JWST communities tend to have higher institution diversity and lower country diversity, while smaller ones exhibit higher country diversity but lower institution diversity.

In terms of how the investigator network shapes and integrates scientific communities in JWST astronomy, three further points emerge. Firstly, we find that it is possible to analyse the investigator graph on a notional dimension of scale, with the three communities on the left-hand side engaged in AU-scale research, and those on the right-hand side in kpc-Gpc scale research. The largest community in the middle of the graph (community #3, stars and stellar systems) is the most diverse in terms of STScI science categories; it is also visually connected to both AU-scale and kpc-Gpc scale communities on the investigator graph (Figure 2). While most communities do reconstructionist work, there is more inventory-oriented work in the left-hand communities compared to the right-hand side. The left- and right-hand sides of the network also differ in terms of instrument-orientation (e.g., NIRISS on the left-hand side) and techniques (e.g., coronagraphy). Secondly, it is difficult to identify a salient variable that accounts for vertical dispersion of the investigator graph. Several variables were attempted, such as whether the communities are single target- vs. ensemble-oriented, rarity of the target, and whether scheduling risk is likely given the nature of the community. None of these appear to discriminate on a notional vertical axis; further work is still required here. Thirdly, there are a small number of key bridging nodes that appear to connect AU-scale and kpc-Gpc scale communities. Their removal does not alter modularity or global cohesion, demonstrating that cross-domain connectivity between smaller scale and larger scale science is redundant (through multiple alternative paths) and distributed rather than mediated by a small set of critical brokers.

*Reflections and implications for policy*

Unlike social network analysis of astronomy communities examining the 'output-side' of space research based on author- or citation-networks (Espinosa-Rada et al., 2024; Espinosa-Rada, 2026; Heidler, 2011), the current study uses the 'input-side' of program-level data. This limits implications to the formation and emergence of networks centred on programs accepted at one specific telescope, rather than the implications of those networks and communities to the longer-term impact of discoveries and findings. Despite this shortcoming, this focus is useful for several reasons. Firstly, the intense interest and competition to access and use platforms such as JWST (Rao, 2024) means that collaboration through multiple networks has become a key strategy – an access strategy - for many astronomers to participate in accepted programs. Secondly, as a direct consequence of access policy, i.e., the work of time allocation committees and the processes they develop and execute (e.g., Reid, 2014; Zasowski et al., 2025), large scale international scientific networks emerge. Understanding these networks at the aggregate level can be important when reflecting on how the broader institutional system for access and time allocation achieves strategic objectives. It may not only be about discoveries, publications and citations, but also about the emergence and strengthening of communities at the investigator level that will sustain the field for many decades, even after the decommissioning of a platform. Thirdly, it has been noted how bias can exist in the process of access and time allocation to telescopes. This may be in the form of gender bias (e.g., Reid, 2014) but also in terms of bias towards countries and institutions that hold dominant or leadership positions due to funding sources and expertise (Malkov et al., 2011; Stahlman, 2023), or because of geo-political forces and alignment (Williams, 2026). Understanding the nature of emerging networks on space telescopes can contribute to debates on bias, fairness and openness in astronomy research.

The current findings may have implications for research institutions hosting the astronomy community and the importance (or not) of poly-centricity with the institution network. We find that a small number of hub institutions are key to overall network collaboration. These are more likely to be relevant to discussions on centralization than the small number of countries. In other words, while the debate about the scientific divide between countries is important (Stahlman, 2023), the institution divide may carry more weight when it comes to access and community building and may therefore need more attention and consideration. This can have implications for the long-term strategy and development of both core and peripheral institutions. Firstly, capacity building for non-core institutions can reduce dependency on core institutions and create alternative pathways for those institutions to develop successful proposals. Such capacity building in non-core institutions might lead to greater geographic distribution of expertise. This could take the form of training, rotations and secondment of human capital, and inclusion in data analysis infrastructure. Secondly, diversifying core institutions internationally (i.e., internationalization) could complement capacity building at the periphery. It can help to reduce dependence on a small number of hubs by transferring certain capabilities to non-core institutions in exchange for reduced coordination burden and increased network resilience; this is not necessarily a zero-sum game. Thirdly, such diversification of roles or access can be based on institution position rather than country position; in the case of JWST this could mean greater institutional diversification and decentralization within the US for instance.

The investigator network analysis has implications for policy at the individual level. The investigator network is found to be the most decentralized and heterarchical (Stark, 2009), reflecting the needs of science. Findings suggest that policies for access and time allocation have worked efficiently as far as network emergence is concerned, even though network emergence is not one of the stated goals of access policy. We find that there does not seem to be a small set of "superstar" researchers at the individual level in the JWST network when organized in this manner. Each community (Figure 2 and Table 1) does have certain local nodes assuming a prominent community position; but at the aggregate level, the network is decentralized and sparse. Policy also extends to open access and rewarding non-exclusivity, encouraging public repositories, transparency in time allocation processes, training, support and community engagement prior to each cycle by STScI and lead institutions. Collectively, this constellation of institutional arrangements appears to have resulted in an aggregate network at the investigator level that is consistent with the strategic goals of the telescope (i.e., consistent with scientific goals, Rieke et al., 2005) while allowing a broad base for participation.

It would have been difficult for lead institutions to predict the nature of the networks that have emerged; their focus being on science, time allocation, managing collaboration and competition, and assisting investigators in program execution, among many other priorities. While the emerging networks at country and institution level are in line with expectations, the one at investigator level is possibly welcome news because of its decentralization, alignment of communities to STScI scientific categories, and the institutionalized redundancy that appears to have been built into the network as it has grown. The emergence of a very small number of bridge-like nodes with low clustering coefficients that connect disparate communities could be a concern for network integrity; a potential weak spot and vulnerability to network resilience. We find that removal of these nodes does not impact the aggregate network characteristics; it does not alter modularity or global cohesion. This demonstrates that cross-domain connectivity between AU-scale and kpc-Gpc scale science is redundant and distributed rather than mediated by a small set of critical brokers. This is further evidence of a robust, institutionalized integration between different types of astronomy conducted on JWST, a feature that has emerged because of a broader institutionalization of governance relating to access. It is also consistent with the Astro2020 Decadal Survey (NASEM, 2021) that stresses the importance of connecting phenomena across difference scales and integrating scientific communities.

*Limitations and future research avenues*

The current study comes with several limitations and paves the way for future research involving emerging networks at the 'input-side' of astronomy. Firstly, we do not associate the different networks or parts of networks with discovery, publications and impact (i.e., 'output-side'). Secondly, we do not link investigator nodes to other networks on other telescopes, which many astronomers also seek to use. Thirdly, we do not show the evolution of communities over time or examine features such as their stability, longevity, growth and decline. Fourthly, we do not consider networks for non-GO programs, such as Director's Discretionary Time (DDT) or Guaranteed Time Observations (GTO). Fifthly, it was not possible to analyse networks involved in rejected proposals or compare rejected proposal networks with accepted proposal networks. Sixthly, the current work does not examine node characteristics such as age, tenure, gender, track record, or the extent to which individuals were PI or Co-I.

Future work can address these limitations while also seeking to answer new questions linked to the findings presented here. These include: (1) exploring factors that determine distribution of nodes on the y-axis; (2) incorporating new cycles into the dataset as they become available in future years; (3) using qualitative data and case methods to track the origins and outcomes of network formation, including how decisions are made on network participation or exclusion; (4) conducting a deep dive into the communities identified here to explore lower level network structures, attributes and access strategies.

## 4. Methods

### *Data collection and preparation*

The following steps were conducted to collect and process data.

1. All investigators (PIs and Co-Is) for the first five Cycles of General Observer (GO) programs were collected from STScI program information (https://www.stsci.edu/jwst/science-execution/program-information) based on searches using the program IDs on PDF lists of GO programs published by STScI under the approved GO programs (https://www.stsci.edu/jwst/science-execution/approved-programs).
2. Institution names were collected from the investigator information shown next to each name for each program under STScI program information. Country of institution was allocated by hand, and by using web searches in situations where the institution country was not known or was potentially ambiguous.
3. Checks for duplicates in investigator names was conducted and resolved.
4. The result of steps 1 – 3 was a single file of all program IDs with associated investigators, their institutions, and their institution country as separate columns. This file had 20,615 rows, more than the number of programs, given the many-to-many relationship between investigators and programs. An investigator can be on more than one program. A program has more than one investigator. We refer to this as an exploded program file.
5. Three Python scripts were used to convert the exploded program file to three edge files that could be imported into the SNA software, Gephi 0.10 running on Windows 11. One edge file was created at country level, one at institution level, and one at investigator level. The edge files contained three columns: source, target, and weight. The source and target columns contained the labels for two nodes (i.e., two countries, two institutions, or two investigators respectively) that were connected through an investigator's participation on a program, and the weight column contained the number of unique instances in which they were connected.
6. On importing the edge files separately into Gephi as undirected networks, node files (lists of nodes) were created as part of the import process.

*Analysis*

The following steps were conducted during data analysis.

1. The ForceAtlas2 layout algorithm was chosen for analysis at country, institution and investigator level. ForceAtlas2 was preferred over alternatives such as Fruchterman-Reingold and Yifan Hu as it is known to work well on large real-world datasets (Jacomy et al., 2014). It pulls densely connected groups together while pushing weakly connected groups apart, important to explore potential communities in the JWST investigator base. Nodes with a high degree (high numbers of connections to other nodes), tend to move toward the centre of the graph, easing interpretation on questions relating to centralization in the network. ForceAtlas2 also is reported to be more optimal for identifying bridging nodes.
2. Following the creation of all three graphs, network statistics were calculated within Gephi, and Excel was used to calculate degree centralization ($C_D$) using Freeman's formula (Freeman, 1979) based on a node export file from Gephi containing degree values for each node. Table 2 below provides definitions and measurements for the network coefficients calculated.

| Variable | Definition |
|---|---|
| Nodes (N) | Number of actors (countries, institutions, or investigators) in the network. |
| Edges (E) | Number of relationships or links between actors. |
| Average Degree (k-bar) | Average number of ties per node. |
| Modularity (Q) | Degree to which the network is divided into distinct communities. Higher values indicate stronger community structure. Values can range from -1 to 1 where -1 indicates a community structure worse than random, 0 a community structure no better than random, >0.5 clear strong community structure. |
| Modularity class | Integer label assigned by Gephi to a community: 1, 2, 3…. |
| Average Clustering Coefficient (C-bar) | Average tendency of a node's neighbours to be connected to one another. Theoretical range is between 0 and 1 where -> 0 implies very little local cohesion and -> 1 implies very strong clustering. |
| Density (D) | Proportion of all possible ties that exist. Theoretical range is between 0 and 1 where -> 0 implies a very sparse network and 1 implies a complete graph with all nodes connected to each other. |
| Degree Centralization ($C_D$) | Measures how strongly the network is organized around the most connected node. Values range from 0 (decentralized) to 1 (star network). |

*Table 2. Definition of social network coefficients*

3. For country and institution graphs, appearance was determined using two node attributes and based on ranking of node degree: (1) colour of node: a white to green rendering (green being higher), and (1) node size. The output is shown in Figure 1 above.
4. For the investigator level graph, the modularity class was applied as the partition attribute for node colour (Figure 2 above). The modularity class is an integer number used to identify and discriminate between communities; this was used when exploring the graph and linking to STScI science categories. An ID is reported here for the top 10 communities ranked by size (Table 1).
5. After calculating the modularity class, the node file for investigators was exported. This allowed the groups of individual investigators per community to be identified, selected, and interpreted. Copilot was used to aid interpretation of the research interests and nature of science performed by each community list, along with reference to selected individuals' public institution webpages and IAU member directory pages where available, showing research interests and areas of expertise and specialisation. Copilot was an efficient way of initially summarizing the central themes in research work conducted by large lists of individual scientists and selected cross referencing to institutional webpages and the IAU directory provided validation in the selected cases examined.
6. Cross referencing to STScI science categories was done by creating a separate file showing the distribution of participation in STScI science categories based on the program ID(s) linked to individual investigators. A manual inspection and grouping were then performed on the fractional link to STScI science categories, for each community. This generated the % matches in Table 1. An Inverse Simpson value was created for each community based on the STScI science categories linked to each community. The Inverse Simpson Index is a diversity measure that expresses the effective number of equally abundant categories in a dataset, with a theoretical range between 1 (only one category in a dataset) and S, the total number of categories possible in a dataset.
7. Given that institution and country of institution are known for each node in each community, a distribution of (1) countries and (2) institutions was created per community, Inverse Simpson Indices to be calculated for each community. These are shown in Table 1. A bivariate correlation between community size (node count) and the three diversity values was performed, along with an overlapping drop line graph of normalized values for the three indices (n=10), using Stata 19. This is shown in Figure 3.
8. Interpretation of the investigator graph in terms of scale of targets, instrumentation use, techniques, and emphasis on inventory science vs reconstruction science, was done using Copilot queries based on large lists of investigators for each community (Table 1, and Figure 4). These lists ranged from n=139 (community #10) to n=1084 (community #1) and were augmented with random searches on the research work of individuals within the communities. Examples of GO programs wholly or principally staffed by nodes from each community were also examined manually to validate the results from Copilot searches.
9. A bridging node analysis was performed through visual inspection of the investigator graph in Gephi, as well as inspection of clustering coefficients, to identify nodes connecting disparate communities. Identification of middle ground links (useful because of the repulsion vs gravity effects in ForceAtlas2) allowed capturing of names of individuals (not reported), and snapshots of graphs in Gephi. Removal of groups of bridging nodes (e.g., with clustering coefficients less than 0.25 and less than 0.20) was performed before re-calculating network statistics.